\documentclass[12pt]{article}
\usepackage{amsmath}
\usepackage{epsfig}
\newcommand{\bfsig}{{\mbox{\boldmath$\sigma$}}}

\def\la{\mathrel{\mathpalette\fun <}}

\def\fun#1#2{\lower3.6pt\vbox{\baselineskip0pt\lineskip.9pt
\ialign{$\mathsurround=0pt#1\hfil##\hfil$\crcr#2\crcr\sim\crcr}}}

\begin{document}
\title{Three nucleon forces in nuclear matter in the QCD sum rules
}
\author{E. G. Drukarev, M. G. Ryskin, V. A. Sadovnikova\\
{\em National Research Center "Kurchatov Institute"}\\
{\em B. P. Konstantinov Petersburg Nuclear Physics Institute}\\
{\em Gatchina, St. Petersburg 188300, Russia}}
\date{}
\maketitle

\begin{abstract}
We calculate the single-particle nucleon characteristics in symmetric nuclear matter with inclusion of the $3N$ interactions. The contributions of the $3N$ forces to nucleon self energies are expressed in terms of the nonlocal scalar condensate ($d=3$) and of the configuration of the four-quark condensates ($d=6$) in which two daiquiri operators act on two different nucleons of the matter. The most important part of the contribution of the four-quark condensate is calculated in a model-independent way. We employed a relativistic quark model of nucleon for calculation of the rest part. The density dependence of the vector and scalar nucleon self energies and of the single-particle potential energy are obtained.
Estimations on contributions of the $4N$ forces to the nucleon self energies are made.
\end{abstract}

\section{Introduction}

The three nucleon ($3N$) interactions is an important ingredient of the nuclear matter theory. Various approaches to calculation of the $3N$ forces are considered nowadays -- see, e.g. \cite{1},\cite{2}, \cite{3} and references therein. These calculations require certain phenomenological parameters. In QCD sum rules approach employed in the present paper the most important contribution of the $3N$ forces is expressed in terms of observables.

We calculate the contribution of the $3N$ forces to the single particle self energies and to the potential energy of the nucleon in symmetric nuclear matter employing the finite density QCD sum rules.
The idea of QCD sum rules in vacuum was to express the characteristics of the observed hadrons in terms of the
vacuum expectation values of the QCD operators, often referred to as the condensates. The method was suggested
for calculation of the meson characteristics \cite{3a}. Later it was used for nucleons \cite{4a}, see also \cite{4}. The approach was extended for the case of finite baryon density (see \cite{4b} and review \cite{5} for references).

The QCD sum rules method in vacuum is based on the dispersion relations for the
function $\Pi_0(q^2)=i\int d^4xe^{i(q\cdot x)} \langle 0|T[j(x)\bar j(0)]|0 \rangle,$ describing the propagation of the system with the four-momentum $q$ which carries the quantum
numbers of the hadron (this is the proton in our case). Here $j(x)$ is the local operator with the proton quantum numbers, often referred to as the "current". The form of $j(x)$ is not unique. We employ the Ioffe current $j(x)=(u^T_a(x)C\gamma_{\mu}u_b(x))\gamma_5 \gamma^{\mu}d_c(x) \varepsilon^{abc}$ ($u(x)$ and $d(x)$ are the quark operators, $a,b,c$ are the color indices, $C$ stands for the charge conjugation) often used in the QCD sum rules calculations \cite{6}.  The function $\Pi_0(q^2)$ is usually referred to as the "polarization operator".

The dispersion relations are considered at large and negative values of $q^2$. Due to the asymptotic freedom of QCD the polarization operator $\Pi_0(q^2)$ can be calculated as a power series of $1/q^2$ and of the QCD coupling $\alpha_s$.
In framework of the Operator Product Expansion (OPE) the large distances contribution is included in terms of the vacuum expectation values of the local operators, such as the scalar quark condensate $\langle 0|\bar q(0)q(0)|0\rangle$, the gluon condensate $\langle 0|\alpha_s/\pi G^a_{\mu\nu}(0)G^a_{\mu\nu}(0)|0\rangle,$ etc. The nonzero imaginary part of the polarization operator is due to the singularities of the function $\Pi_0(q^2)$ corresponding to real states with the quantum numbers of the proton. In the standard approach to description of the spectrum of the function $\Pi_0(q^2)$ the lowest lying pole (the proton) is written explicitly while the higher excited states are treated approximately. This is called the "pole+continuum" model. The position of the nucleon pole $m$, the residue of the latter $\lambda^2$ and the effective continuum threshold $W^2$ are the unknowns of the QCD sum rules equations. Thus the
proton characteristics are expressed in terms of the QCD condensates.

The finite density QCD sum rules are expected to present the characteristics of the proton placed on the Fermi surface
of nuclear matter in terms of the {\em in-medium} expectation values of the QCD operators. The left hand sides of the sum rules are obtained as power series in $1/q^2$ of the in-medium polarization operator $\Pi(q)=i\int d^4xe^{i(q\cdot x)} \langle M|T[j(x)\bar j(0)]|M\rangle$ with  $|M\rangle$ the ground state of nuclear matter. The higher terms of expansion contain the condensates of the higher dimensions. Since we expect the expansion of the $1/q^2$ series to converge,
we expect the terms containing the condensates of the lowest dimension to be the most important ones. The vector and scalar quark condensates
\begin{equation}
v(\rho)=\langle M|\sum_i\bar q^i\gamma_0q^i|M\rangle; \quad
\kappa(\rho)=\langle M|\sum_i\bar q^iq^i|M\rangle
\label{X0}
\end{equation}
have the lowest dimension $d=3$.
The sums are carried out over the quark flavors,
$\rho$ is the nuclear matter density. Note that the vector condensate (its expression is written in the rest frame of the matter) vanishes in vacuum. The
"pole+continuum" model is used for the description of the spectrum of polarization operator. Thus characteristics of the proton pole  which we call {\em the probe proton} are expressed in terms of the in-medium condensates. We shall see that these characteristics are the vector and scalar nucleon self energies.

The condensates of the lowest dimension $ d=3$ contributing to the polarization operator are the vector and scalar expectation values expressed by Eq.~(\ref{X0}).
They correspond to exchange of vector and scalar fields between the nucleon and the matter. The contributions of dimension $d=4$ are provided by the gluon condensate and by the nonlocality of the vector condensate. We do not discuss them here since they are numerically small. However, we include them in the calculations. Among the condensates of the higher dimension the four-quark expectation values $\langle M|q^a_{\alpha}\bar q^b_{\beta}q^c_{\gamma}\bar q^d_{\delta}|M\rangle$ with $d=6$ are the most important ones.

Until now the QCD sum rules have been studied in the gas approximation where each condensate was presented as the sum of the vacuum value and nucleon density times expectation value of the same operator in the nucleon. In other words, the polarization operator interacted with each of the nucleons separately. This provided the characteristics of the  probe proton corresponding to the lowest pole of polarization operator with inclusion of only the $2N$ forces between the nucleons of the matter. The vector condensate $v(\rho)=n_v\rho$, where $n_v=3$ is the number of the valence quarks in nucleon, is exactly proportional to the nucleon density due to conservation of the vector current.
In the gas approximation the scalar condensate can be written as
\begin{equation}
\kappa(\rho)=\kappa(0)+\kappa_N\rho; \quad \kappa_N=\langle N|\sum_i\bar q(0)q(0)|N\rangle.
\label{X1}
\end{equation}
The nucleon matrix element for the nucleon at rest can be presented as
\begin{equation}
\kappa_N=\frac{2\sigma_N}{m_u+m_d},
 \label{3b}
\end{equation}
with $\sigma_N$ the pion-nucleon sigma term which can be related to observables \cite{8}. In Eq.~(\ref{3b}) $m_{u,d}$ are the masses of light quarks.

In the gas approximation we included the contribution of the four-quark condensate in which all quark operators act on the same nucleon of the matter. Calculation of the four-quark expectation values $\langle N|q^a_{\alpha}\bar q^b_{\beta}q^c_{\gamma}\bar q^d_{\delta}|N\rangle$ required model assumptions on the quark structure of nucleon. We employed the relativistic model suggested in \cite{9} in the version described in \cite{5}. The Feynman diagrams providing the main contribution to the nucleon parameters in the gas approximation are shown in Fig.~1.

 To find the influence of the $3N$ forces on the self energies of the probe proton, we must include the in-medium QCD condensates beyond the gas approximation, limiting ourselves, however to $2N$ forces between the nucleons of the matter. Following the previous analysis, we must calculate the scalar condensate beyond the gas approximation. We must include also the configurations of the four-quark condensates in which two pairs of quarks act on two different nucleons of the matter-see Fig.~2.

The scalar condensate can be written as $\kappa(\rho)=\kappa(0)+\kappa_N\rho+S(\rho)$, with the function $S(\rho)$ caused by the interactions of the nucleons of the matter. It is known \cite{4b},\cite{5} that the function $S(\rho)$ is dominated by the pion cloud created by interacting nucleons. Thus it is determined by the expectation value $\langle \pi|\bar qq|\pi\rangle=2m_{\pi}^2/(m_u+m_d)$. After the earlier calculations of the pion contribution to $S(\rho)$ \cite{6}, \cite{6a} more rigorous analysis was carried out in \cite{7} in framework of the Chiral Perturbation Theory (CHPT). In \cite{7} the function $S(\rho)$ is obtained including the $2N$ and $3N$ forces between the nucleons of the matter. However, it is possible to extract the contribution corresponding to inclusion of only the $2N$ forces.

Now we must include the contribution to the four quark condensate $\langle M|\bar q\Gamma_Aq\bar q\Gamma_Bq|M\rangle$ in which two pairs of quarks act on two different nucleons of the matter while the latter are at the same space-time point. These two nucleons suffer strong interaction (the "repulsive core" of traditional nuclear physics). The short range nucleon interactions is usually treated in the terms of quarks and gluons nowadays (see, e.g. the review \cite{14a}). It was found in \cite{14b} that this interaction is due mostly to the Pauli principle for the quarks in the six-quark system. These ideas have been realized explicitly in the nucleon-nucleon potential suggested in \cite{14c}. In our calculations the form of the current $j(x)$ insures the needed antisymmetrization.The latter effectively takes into account  the main part of the nucleon short range interactions and we neglect the remaining part.

 The most important scalar-scalar,
vector-scalar and vector-vector contributions to the four-quark condensates can be obtained in a model-independent way. The scalar-scalar and  vector-scalar condensates can be expressed in terms of the sigma term.
Calculation of the other expectation values requires model assumptions on the quark structure of nucleons. We employ the same relativistic quark model \cite{9} which we used in the calculations in gas approximation.

Thus the $3N$ interactions can be expressed mainly in terms of the pion-nucleon sigma term. We accept the conventional value $\sigma_N=45$ MeV \cite{8} which is in agreement with the latest result  $\sigma_N\approx 44$ MeV \cite{8a}. All calculations are carried out for symmetric nuclear matter, i.e. for the matter with equal number of protons and neutrons.

At the phenomenological value of saturation density $\rho_0=$0.16 fm$^{-3}$ the $3N$ forces diminish the value of the effective mass $m^*$ by about 25 MeV and subtract 37 MeV from the value of the vector self energy $\Sigma_V$. The $4N$ interactions are expected to add about 100 MeV to $m^*$, providing only small change of $\Sigma_V$.

In Sec.~2 we recall the main features of the finite density QCD sum rules. In Sec.~3 we
calculate the contribution of the nonlinear scalar condensate to the polarization operator. In Sec.~4 we
obtain the contribution of the four-quark condensates. In Sec.~5 we find the contributions of the $3N$  interactions to the single particle self energies and to the potential energy for the symmetric nuclear matter.
We summarize in Sec.~6.

\section{Finite density QCD sum rules}

The details of the finite density QCD sum rules approach are described in ~\cite{5}. Here we just recall the main points. We introduce the 4-vector $P=(m,{\bf 0})$ with $m$ being the vacuum value of the nucleon mass (we neglect the neutron-proton mass splitting). The polarization operator $\Pi(q,P)=\hat q\Pi^q(q,P)+\hat P\Pi^P(q, P)+I\Pi^I(q,P)$ depends on the four-vectors $q$ and $P$. To separate the singularities connected with the probe proton from those connected with the matter itself we fix the value of $s=(q+P)^2$ putting $s=4m^2$.
The polarization operator  takes the form
\begin{equation}
\Pi(q^2,s)=\hat q\Pi^q(q^2,s)+\hat P\Pi^P(q^2,s)+I\Pi^I(q^2,s).
\label{10r}
\end{equation}
The dispersion relations are written for each component $\Pi^i(q^2) \,(i=q,P,I)$.\\
The imaginary parts on the right hand side  of the dispersion relations describe the physical states with the proton quantum numbers and are approximated by the "pole+continuum" model
\cite{4} in which the lowest lying pole is written down exactly,
while the higher states are described by continuum with the discontinuity corresponding to the OPE terms.
Thus
$$\frac{\mbox{Im}\Pi^i(k^2,s)}{\pi}=\xi^i\lambda_{Nm}^2(s)\delta(k^2-m_m^2(s))+\frac{\Delta \Pi^{iOPE}(k^2,s)}{\pi}\theta(k^2-W_m^2(s)).$$
Here  $\xi^q=1$, $\xi^I=m^{*}$, $\xi^P=-\Sigma_V/m$, while $m$ is the vacuum nucleon mass. The vector and scalar self-energies $\Sigma_V$ and $\Sigma_S=m^{*}-m$ and also the parameters $\lambda_{Nm}^2$ and $W_m^2$ are the unknowns of the SR equations. The position of the pole corresponding to the probe proton can be expressed in terms of $\Sigma_V$ and $\Sigma_S$; in linear approximation $m_m=m+\Sigma_S+\Sigma_V$.
The left hand side  of the dispersion relations can be written as
\begin{equation}
\Pi^{q~OPE}(q^2)=\sum_{n=0}A_n(q^2); \quad \Pi^{I~OPE}(q^2)=\sum_{n=3} B_n(q^2); \quad \Pi^{P~OPE}(q^2)=\sum_{n=3} C_n(q^2),
\label{14}
\end{equation}
where the lower index $n$ is the dimension of the corresponding QCD condensate.
($A_0$ stands for the free three-quark loop). The terms with $n=3$ correspond to the scalar and vector two-quark condensates contribution, while $n=6$  for  the four-quark condensates.

To suppress the heavier state contribution the Borel transform is used. It converts a function of $q^2$ to the Borel transformed function depending on the squared Borel mass $M^2$. The Borel transformed OPE terms $A_n(q^2)$, $B_n(q^2)$ and $C_n(q^2)$ of (\ref{14}) multiplied by $32\pi^4$ (the factor $32\pi^4$ is introduced in order to deal with the magnitudes of the order of unity) are denoted as $\tilde A_n(M^2,W_m^2)$, $\tilde B_n(M^2, W_m^2)$ and $\tilde C_n(M^2, W_m^2)$ correspondingly. The Borel transformed sum rules take the form
\begin{equation}
{\cal L}^q(M^2, W_m^2)=\Lambda_m; \quad {\cal L}^I(M^2, W_m^2)=m^{*}\Lambda_m; \quad {\cal L}^P(M^2, W_m^2)=-\Sigma_P\Lambda_m,
\label{X3}
\end{equation}
with
\begin{equation}
{\cal L}^q(M^2,W_m^2)=\sum_{n=0}\tilde A_n(M^2,W_m^2); \quad {\cal L}^I(M^2,W_m^2)=\sum_{n=3}\tilde B_n(M^2,W_m^2);
\label{X4}
\end{equation}
$${\cal L}^P(M^2,W_m^2)= \sum_{n=3}\tilde C_n(M^2,W_m^2).$$
In Eq.~(\ref{X3}) $\Lambda_m=\lambda^2_{m}e^{-m_m^2/M^2}$ while $\lambda_m^2=32\pi^4\lambda_{Nm}^2$.

Now we try to establish connection of our results with the picture based on the three-nucleon interactions. In our approach the parameters of the probe nucleon:
the effective mass $m^*$ and the vector self energy $\Sigma_V$ are expressed in terms of the lowest dimension terms of the polarization operator in which the weakly correlated quarks are exchanged between the polarization operator and the matter. The scalar and vector nucleon self energies can be viewed as due to exchange of mesons (systems of strongly correlated quarks) between the probe nucleon and the nucleons of the matter. Thus the exchange by systems of strongly correlated quarks between the probe nucleon and the nucleons of the matter is expressed through those of weakly correlated quarks with the same quantum numbers between the current and the matter.

Start with the case of $2N$ forces. Including only condensates of the  lowest dimension $d=3$ we find the polarization operator to be determined by
the vector and scalar condensates. This corresponds to the lowest order exchange by vector and scalar mesons between the probe nucleon and those of the matter. This picture is in agreement with that provided by relativistic nuclear physics \cite{12a}. However the four quark condensates ($d=6$) also appear to be numerically important. The pairs of quarks can carry various combinations of the quantum numbers. These can be the scalar-scalar ($SS$), vector-vector ($VV$), pseudovector-pseudovector ($PP$), axial-axial ($AA$), tensor-tensor ($TT$),vector-scalar ($VS$) and axial-tensor ($AT$) combinations \cite{5}, \cite{X}. The probe nucleon exchanges by the mesons with the corresponding quantum numbers with one of the nucleons of the matter-see Fig.~3.

Inclusion of the $3N$ forces leads to additional diagrams. The contribution to the polarization operator in the lowest dimension $d=3$ corresponds to exchange by the scalar meson between the probe nucleon and the pion field created by {\em two} nucleons of the matter. The four quark condensate with two pairs of quarks acting on two different nucleons corresponds to exchange by mesons between the probe nucleon and two in-medium nucleons. The quantum numbers of the mesons are the same as in the case of $2N$ forces. These additional diagrams are shown in Fig.~4.

Note that the density dependence of the nucleon self energies is not linear even in the gas approximation. For the nucleon with four-momentum $q$ (see, e.g.\cite{12a})
$$m^*=\frac{m+\Sigma_s}{1-\Sigma_q}; \quad \Sigma_V=\frac{\Sigma_v}{1-\Sigma_q}.$$
In the gas approximation the self energies $\Sigma_s$, $\Sigma_v$ and $\Sigma_q$ are proportional to $\rho$.
After inclusion of the $3N$ interactions they contain also the terms proportional to $\rho^2$. The self energies obtained in the sum rules analysis also have this feature since it follows from Eqs.~(\ref{X3},\ref{X4}) that
\begin{equation}
\frac{{\cal L}^I(M^2, W_m^2)}{{\cal L}^q(M^2, W_m^2)}=m^{*};\quad
\frac{{\cal L}^P(M^2, W_m^2)}{{\cal L}^q(M^2, W_m^2)}=-\Sigma_V/m\,.
\label{57}
\end{equation}

\section{Contribution of the nonlinear scalar condensate to the polarization operator}

Recall the scalar condensate enters only the coefficient $\tilde B_3$ in Eq.~({\ref{X4})
\begin{equation}
\tilde B_3=-4\pi^2M^4E_1(W_m^2/M^2 )r_{B3}\kappa(\rho).
\label{600}
\end{equation}
The function
$E_1(x)=1-(1+x)e^{-x}$
with $x=W_m^2/M^2$ includes the contribution of continuum, the factor
$$r_{Xn}=1+\frac{\alpha_s}{\pi}c_{Xn},$$  accounts for the QCD coupling $\alpha_s$  correction
with the coefficients $c_{Xn}$ presented in \cite{5} includes the logarithmic corrections.
Following \cite{7} we present
\begin{equation}
\kappa(\rho)=\kappa(0)+\frac{2\sigma^{eff}_N(\rho)}{m_u+m_d},
 \label{X5}
\end{equation}
and $\sigma_N^{eff}(\rho)=\sigma_N$ if the interactions between the nucleons of the matter are neglected
(here we omit small corrections caused by the kinetic energy of the nucleons).

The difference $\sigma_N^{eff}(\rho)-\sigma_N$ was calculated in \cite{7} with inclusion of the $2N$ and $3N$ forces between the nucleons of the matter. The authors calculated the one pion exchange including the iterations and the two pion
exchange with only the nucleons in the intermediate states and with virtual $\Delta$ isobar excitations. They included also the
contact $\pi\pi NN$ vertices which are specific for the CHPT. The contributions of two pion
exchange with only the nucleons in the intermediate states contain logarithmic divergence at large values of the pion momenta. This requires introduction of the cutoff $\mu$ (it is denoted as $\lambda$ in \cite{7}).

Fortunately it is possible to separate the contribution of the $2N$ forces in the results obtained in \cite{7}.
One can write $\sigma_N^{eff}(\rho)-\sigma_N=m_{\pi}^2\sum_iD_i$ with $D_i$ the partial contributions expressed by Eqs.~(4,6,7,11,14, 17,18,20)
of \cite{7}. Numerically largest contributions are provided by the iterated one pion exchange and the two pion exchanges with $\Delta$ isobars in intermediate states. However, they come with opposite signs, canceling each other to large extent. The final result depends on the value of the high momentum cutoff $\mu$. In Fig.~5 we present the function $\sigma_N^{eff}(\rho)$ calculated for the values $\mu=600$ MeV, 882 MeV and 1200 MeV.

\section{Contribution of the four-quark condensates to polarization operator}

The four quarks condensates are included already in the gas approximation.
This provides
\begin{equation}
\tilde A_6=\frac{4}{3}a^2-8\pi^2A_g\rho; \quad \tilde B_6=4\pi^2\frac{s-m^2}{m}aB_g\rho; \quad \tilde C_6=4\pi^2 \frac{s-m^2}{m}aC_g\rho,
\label{19}
\end{equation}
with $s=4m^2$, $a=-(2\pi)^2\langle 0|\bar qq|0\rangle \approx 0.55 $GeV$^3$. For symmetric matter the dimensionless parameters $A_g=-0.10$, $B_g=1.80$ and $C_g=-0.88$ where obtained in \cite{X}.

Now we add the configurations in which two pairs of quark operators act on two different nucleons of the matter.
\subsection{General expressions}

The four-quark condensates are described by the matrix elements
$\langle M|q^a_{\alpha}\bar q^b_{\beta}q^c_{\gamma}\bar q^d_{\delta}|M\rangle$. (Here we do not specify the quark flavors). For each of the products of the quark operators we can write
\begin{equation}
q^a_{\alpha}\bar q^b_{\beta}=-\frac{1}{12}\sum_X\bar q\Gamma_Xq(\Gamma^X)_{\alpha\beta}\delta_{ba}-
\frac{1}{8}\sum_{X,a}\bar q\Gamma_X\lambda^{\alpha}q(\Gamma^X)_{\alpha\beta}\lambda^{\alpha}_{ba},
\label{5}
\end{equation}
with $\lambda^{\alpha}$ standing for $SU(3)$ color Gell-Mann matrices.
The sixteen basic $4\times 4$ matrices $\Gamma_{X,Y}$ acting on the Lorentz indices of the quark operators are
\begin{equation}
 \Gamma_X=I; \quad \Gamma_X=\gamma_5; \quad \Gamma_X=\gamma_{\mu}; \quad \Gamma_X=\gamma_{\mu}{\gamma_5};
\label{22a}
\end{equation}
$$\Gamma_X=\frac{i}{2}(\gamma_{\mu}\gamma_{\nu}-\gamma_{\nu}\gamma_{\mu})=\sigma_{\mu\nu} (\mu>\nu),$$
describing the scalar ($S$), pseudoscalar ($P$), vector ($V$), axial ($A$) and tensor ($T$) cases correspondingly.
We consider the configuration in which the diquark operators $\bar q\Gamma_Xq$ and $\bar q\Gamma_Yq$ act on two different nucleons of the matter.
The contribution of such configurations to the polarization operator is a linear combination of the terms
\begin{equation}
{\cal M}_{XY}=\frac{\eta_{aba'b'}}{16}\langle {\cal N}|\bar q^a\Gamma_Xq^{a'}\bar q^{b}\Gamma_Yq^{b'}|{\cal N}\rangle\rho^2\Gamma^X\Gamma^Y,
\label{6}
\end{equation}
with $|{\cal N}\rangle$ the two-nucleon state.
The factor
\begin{equation}
\eta_{aba'b'}=\delta_{aa'}\delta_{bb'}-\delta_{ab'}\delta_{ba'}
\label{6a}
\end{equation}
shows that the quark operators have the same color in each of the diquark operators $\bar q\Gamma_{X}q$ and $\bar q\Gamma_{Y}q$. However the quark colors in these two diquark operators can not be the same. Note that one can present
$$\eta_{aba'b'}=\frac{2}{3}\delta_{aa'}\delta_{bb'}-\frac{1}{2}\sum_{\alpha}\lambda^{\alpha}_{aa'}\lambda^{\alpha}_{bb'}.$$

In the case of four $u$ quarks we find nonvanishing contributions for all combinations with $X=Y$. For the product of two $u$ and two $d$ quarks
nonvanishing contributions come from $X=Y=V$, $X=Y=A$, and also from $X=V$, $Y=S$ and $X=A$, $Y=T$ \cite{X}.

The scalar operator as well as the vector operator in the rest frame do not act on spin variables and the expectation values presented by Eq.~(\ref{6})
take the factorized form being proportional to the product $\langle N_1|\bar q\Gamma^Xq|N_1\rangle\langle N_2|\bar q\Gamma^Yq|N_2\rangle\rho^2$ with the lower indices labeling two in-medium nucleons. The wave function which describes the nucleons $N_{1,2}$ should satisfy the Pauli principle. If both $N_{1,2}$ are protons or neutrons their wave function is spin-asymmetric and the two-nucleon state can have only one spin state with $S=0$. If $N_1$ is a proton while $N_2$ is a neutron (or vise versa) the two nucleons form the state with the isospin projection $T_z=0$. It can be either isotope symmetric or the isotope antisymmetric state. In the former case there is only the spin asymmetric state with $S=0$. In the latter case the state is spin symmetric and we have three spin states with $S=1$. Since in symmetric nuclear matter the proton and neutron densities are $\rho_p=\rho_n=\rho/2$, we find
\begin{equation}
{\cal M}_{XY}=\frac{1}{16}\frac{2}{3}\Big[\frac{1}{4}\langle p|\bar q\Gamma^Xq|p\rangle\langle p|\bar q\Gamma^Y|p\rangle+\frac{1}{4}
\langle n|\bar q\Gamma^Xq|n\rangle\langle n|\bar q\Gamma^Yq|n\rangle
\label{6b}
\end{equation}
$$
+ \langle p|\bar q\Gamma^Xq|p\rangle\langle n|\bar q\Gamma^Yq|n\rangle \Big]\cdot
\frac{\rho^2}{4}\Gamma_X\Gamma_Y.
$$
For the $SS$ condensate $\Gamma_X=\Gamma_Y=I$, for the $VV$ condensate $\Gamma_X=\Gamma_Y=\gamma_0$ (in the rest frame of the matter). For the $VS$ condensate $\Gamma_X=\gamma_0$,$\Gamma_Y=I$. The expectation value of the vector operator $\bar q\gamma_{\mu}q$ is just the number of the valence quarks in the nucleon $n_q$. In the rest frame of the nucleon $\langle N|\bar q\gamma_{\mu}q|N\rangle=n_q\delta_{\mu0}$.
The scalar operator $\bar uu+\bar dd$ averaged over the state of free nucleon can be expressed in terms of the pion-nucleon sigma term $\sigma_{N}$ -- see Eq.~(\ref{3b}).
As we shall see below these expectation values provide the most important contribution
in symmetric nuclear matter.

The pseudoscalar, axial and tensor operators depend on spin variables. The standard assumption is that the quantum numbers of nucleon are carried by the valence quarks. Since the spins of the quarks belonging to two nucleons composing the state $|{\cal N}\rangle$ are correlated, the matrix element presented by Eq.~(\ref{6}) does not take the simple form (\ref{6b}). For example, in the $AA$ channel with $\Gamma_{X}=\gamma_{\mu}\gamma_5$ and $\Gamma_Y=\gamma_{\nu}\gamma_5$ we find
\begin{equation}
{\cal M}_{XY}=\frac{1}{16}\frac{2}{9}\langle {\cal N}|\bar q_v{\bfsig}^{(1)}Iq_v\bar q_v{\bfsig}^{(2)}Iq_v|{\cal N}\rangle\rho^2\Gamma^X\Gamma^Y,
\label{9a}
\end{equation}
instead of Eq.~(\ref{6b}). Here $q_v$ stands for the operators of the valence quarks, $I$ is the unit $4\times 4$ matrix.  The upper indices $1,2$ of the Pauli matrices denotes that they act on the quarks of the nucleons $1$ and $2$ composing the state $|{\cal N}\rangle$. In the case of $4u$ condensate the two $u$ quarks from these nucleons form the state which is space and flavor symmetric and which is also color antisymmetric. Thus they compose the spin-triplet state with the eigen value of the operator ${\bfsig}^{(1)}\cdot{\bfsig}^{(2)}=1$. For $2u2d$ condensate the $u$ and $d$ quarks of each nucleon form the flavor antisymmetric state which is thus the spin singlet with ${\bfsig}^{(1)}\cdot{\bfsig}^{(2)}=-3$.
Similar expressions can be written for the $TT$ and $AT$ cases. There is no contribution from the $PP$ channel since there is no pseudoscalar parameter of a single nucleon.

Now we obtain the contribution of four-quark condensates to polarization operator. Following \cite{5}, \cite{X}
we present
\begin{equation}
(\Pi)_{4q}=(\Pi_0)_{4q}+\frac{1}{q^2}\sum_{X,Y}\Big(\mu^{XY}H_{XY}+\tau^{XY}R_{XY}\Big).
\label{20}
\end{equation}
Here $(\Pi_0)_{4q}$ is the vacuum term, while
\begin{equation}
H_{XY}=\langle M|T_{XY}^{uu}|M\rangle; \quad R_{XY}=\langle M|T_{XY}^{ud}|M\rangle
\label{21}
\end{equation}
with
\begin{equation}
T^{f_1f_2}_{XY}=\bar q^{f_1a}\Gamma_Xq^{f_1a'}\cdot\bar q^{f_2b}\Gamma_Yq^{f_2b'}\eta_{aba'b'},
\label{22}
\end{equation}
where $f_{1,2}$ stand for the quark flavors ($f_{1,2}=u,d$), $a,a',b,b'$ are the color indices, $\eta_{aba'b'}$ is defined by Eq.~(\ref{6a}).
In Eq.~(\ref{20})
\begin{equation}
\mu^{XY}=\frac{\theta_X}{16}Tr(\gamma^{\alpha}\Gamma^{X}\gamma^{\beta}\Gamma^{Y})
\gamma_5\gamma_{\alpha}\hat q\gamma_{\beta}\gamma_5;\quad \tau^{XY}=\frac{\theta_X}{4}Tr(\gamma^{\alpha}\hat q\gamma^{\beta}\Gamma^{X})\gamma_5\gamma_{\alpha}\Gamma ^Y\gamma_{\beta}\gamma_5,
\label{23}
\end{equation}
with $\theta_X=1$ if $\Gamma^X$ has a vector or tensor structure, while $\theta_X=-1$ in other cases.

The trace on the RHS of the first equality in Eq.~(\ref{23}) obtains a nonzero value for all five structures listed in Eq.~(\ref{22a}) if the matrices $\Gamma^X$ and $\Gamma^Y$ belong to the same channel. The trace on the RHS of the second equality does not vanish if both $\Gamma^X$ and $\Gamma^Y$ belong to the vector or axial channel. It has a nonzero value also if $\Gamma^X$ and $\Gamma^Y$ compose the $VS$ and $AT$ combinations.

Operators of the quark fields act on two nucleons of the matter which are in the same space point. If the two nucleons have the same projection of isospin (i.e. both of them are neutrons or protons), they should form the spin-singlet state. They can compose isotriplet and spin singlet state or isosinglet and spin triplet state, if one of them is neutron while the other is proton. Introducing
\begin{equation}
h_{XY}^p=\langle p|\bar u\Gamma_Xu|p\rangle\langle p|\bar u\Gamma_Yu|p\rangle; \quad h_{XY}^n=\langle n|\bar u\Gamma_Xu|n\rangle\langle n|\bar u\Gamma_Yu|n\rangle;
\label{24}
\end{equation}
$$h_{XY}^{pn}=\langle p|\bar u\Gamma_Xu|p\rangle\langle n|\bar u\Gamma_Yu|n\rangle,$$
and $h_{XY}^{np}=h_{XY}^{pn} (p\leftrightarrow n)$ with $\rho_{p,n}$ the proton and neutron densities
($\rho_p+\rho_n=\rho$)
we find for contribution of the $4u$ condensate
\begin{equation}
H_{XY}=\frac{\rho_p^2}{4}h_{XY}^p+\frac{\rho_n^2}{4}h_{XY}^n+\frac{\rho_p\rho_n}{2}(h_{XY}^{pn}+h_{XY}^{np})\,.
\label{25}
\end{equation}
In similar way we find that for $2u2d$ condensate
\begin{equation}
R_{XY}=\frac{\rho_p^2}{4}t_{XY}^p+\frac{\rho_n^2}{4}t_{XY}^n+\frac{\rho_p\rho_n}{2}(t_{XY}^{pn}+t_{XY}^{np}),
\label{26}
\end{equation}
where
\begin{equation}
t_{XY}^p=\langle p|\bar u\Gamma_Xu|p\rangle\langle p|\bar d\Gamma_Yd|p\rangle; \quad t_{XY}^n=\langle n|\bar u\Gamma_Xu|n\rangle\langle n|\bar d\Gamma_Yd|n\rangle;
\label{27}
\end{equation}
$$t_{XY}^{pn}=\langle p|\bar u\Gamma_Xu|p\rangle\langle n|\bar d\Gamma_Yd|n\rangle,$$
and $t_{XY}^{np}=t_{XY}^{pn} (p\leftrightarrow n)$.

 Recall that since the Borel transform of the function $1/q^{2}$ is $-1$ and does not depend on the Borel mass $M^2$, the contributions of the four-quark condensates to the sum rules equations (\ref{X3},\ref{X4}) also do not depend on $M^2$.

\subsection{Vector and scalar condensates}

The contributions of $SS$, $VV$ and $VS$ channels can be obtained in the model-independent way.
Start with contribution of the $VV$ channel in which $\Gamma_X=\Gamma_Y=\gamma_0=\hat P/m$.
The condensates $h_{XY}^N$ and $t_{XY}^N$ are just the products of the numbers of the valence quarks.
For $4u$ condensate Eq.~(\ref{23}) provides
\begin{equation}
\mu^{VV}=-\frac{(Pq)}{m^2}\hat P; \quad (Pq)=\frac{s-m^2-q^2}{2},
\label{28}
\end{equation}
and thus
$$ \mu^{VV}h^n_{VV}=-\frac{(Pq)\hat P}{m^2}; \quad \mu^{VV}h^p_{VV}=-\frac{4(Pq)\hat P}{m^2}; \quad \mu^{VV}h^{np}_{VV}=\mu^{VV}h^{np}_{VV}=-\frac{2(Pq)\hat P}{m^2},
$$
and the contribution to the polarization operator given by Eq.~(23) is for symmetric matter with $\rho_p=\rho_n=\rho/2$
\begin{equation}
\Pi^{3N}_{4q}=-\frac{13}{12q^2}\frac{(Pq)}{m^2}\hat P\rho^2.
\label{29}
\end{equation}
The contribution to the left hand side of the sum rules presented by Eqs.~(\ref{X3},\ref{X4}) is thus
\begin{equation}
\tilde C_6^{3N}=\frac{13}{24}\frac{s-m^2}{m^2}\cdot 32\pi^4\rho^2.
\label{30}
\end{equation}

For $2u2d$ condensate
\begin{equation}
\tau^{VV}=-2m^2\hat q-2(Pq)\hat P,
\label{31}
\end{equation}
and for symmetric matter
\begin{equation}
\Pi^{3N}_{4q}=-\frac{7}{6q^2}\hat q\rho^2-\frac{7}{6q^2}\frac{(Pq)}{m^2}\hat P\rho^2,
\label{32}
\end{equation}
providing
\begin{equation}
\tilde A_6^{3N}=\frac{7}{6}\cdot 32\pi^4\rho^2; \quad \tilde C_6^{3N}=\frac{7}{12}\frac{s-m^2}{m^2}\cdot 32\pi^4\rho^2.
\label{33}
\end{equation}

Turn now to the scalar channel. Here only the $4u$ operator contributes. We find immediately that $\mu_{XX}=-\hat q/2$.
Due to isotopic invariance $\langle n|\bar uu|n\rangle=\langle p|\bar dd|p\rangle$.  Thus employing Eq.~(\ref{26})
we can present
\begin{equation}
\Pi^{3N}_{4q}=-\frac{\hat q}{16q^2}\kappa^2_{N}\rho^2,
\label{34}
\end{equation}
with $\kappa_{N}$ determined by Eq.~(\ref{3b}). Hence
\begin{equation}
\tilde A_6^{3N}=32\pi^4\frac{\kappa_{N}^2\rho^2}{16}.
\label{36}
\end{equation}

The $2u2d$ operator contributes also in the mixed $VS$ channel.
\begin{equation}
\Pi^{3N}_{4q}=-\frac{3}{8q^2}\frac{(Pq)}{m}\left(\kappa_N +\frac{\zeta_p}{9}\right)\rho^2.
\label{39}
\end{equation}
Here
\begin{equation}
\zeta_p=\langle p|\bar uu-\bar dd|p\rangle\,.
\label{40}
\end{equation}
This provides
\begin{equation}
\tilde B_6^{3N}=\frac{3}{8}\frac{s-m^2}{2m}32\pi^4\left(\kappa_N+\frac{\zeta_p}{9}\right)\rho^2.
\label{41}
\end{equation}

The expectation value $\kappa_N$ is related to the pion-nucleon sigma-term $\sigma_{N}$ by Eq.~(\ref{3b}).
We obtain $\kappa_N \approx 8$ for $\sigma_N=45$~MeV.
Simple quark counting predicts $\zeta_N \la 1$. The experimental data on deep inelastic scattering  provides $\zeta_p -1 \approx -0.15$ \cite{12}. Thus the second terms in brackets on the RHS of Eqs.~(\ref{39}) and (\ref{41}) are not important. Hence the contributions calculated in this Subsection are expressed in terms of the observable nucleon sigma term $\sigma_N$. The leading contribution comes from the $SS$ channel.

\subsection{Pseudoscalar, axial and tensor channels}

Inclusion of these channels requires certain assumptions on the nucleon quark structure. The most general one is that the nucleon consists of the valence quarks and the sea of the quark-antiquary pairs in the spin singlet state. The nucleon quantum numbers are carried by the valence quarks. Since the operators $\bar q\Gamma_Xq$ ($X=P,A,T$) are proportional to the quark spin operator, they act only on the valence quark of the nucleon. Assuming that the valence quarks are described by the single particle functions $\psi_q({\bf r})$ we can write
\begin{equation}
\langle N|\bar q\Gamma_Xq|N\rangle=\sum\int d^3r \bar \psi_q({\bf r})\Gamma_X\psi_q({\bf r}); \quad X=P,A,T
\label{42}
\end{equation}
with the sum over the quarks with the flavor $q$. The next assumption is that $\psi_q({\bf r})$ is a solution of the Dirac equation in an external central field. This determines the form of the wave function
\begin{equation}
\psi_q({\bf r})=\left( \begin{array}{c}
f_1(r)\chi \\
i{\bfsig}\cdot {\bf n}f_2(r)\chi\end{array} \right),
\label{43}
\end{equation}
where the functions $f_{1,2}(r)$ do not contain angular dependence, ${\bf n}={\bf r}/{r}$ while $\chi$ is the Pauli spinor.

Now we find immediately that there is no contribution from the pseudoscalar channel since $\int d^3r \bar \psi_q({\bf r})\gamma_5\psi_q({\bf r})=0$. Also, in the axial and tensor channels  $\int d^3r \bar \psi_q({\bf r})\gamma_0\gamma_5\psi_q({\bf r})=\int d^3r \bar \psi_q({\bf r})\sigma_{0i}\psi_q({\bf r})=0$. Thus the contributions of the time component in the axial channel and of the space-time component in the tensor channel vanish as well.

For the space component of axial channel we obtain
\begin{equation}
\langle N|\bar q\gamma_i\gamma_5q|N\rangle=\sum\int d^3r\chi^{*}[{\sigma_i}f_1^2(r)+
({\bfsig}\cdot{\bf n})\sigma_i({\bfsig}\cdot{\bf n})f_2^2(r)]=
\label{44}
\end{equation}
$$\sum\chi^{*}{\sigma_i}\chi\cdot\int d^3r[f_1^2(r)-\frac{2}{3}f_2^2(r)].$$
Thus the matrix element contributing in $AA$ channel takes the form given by Eq.~(\ref{9a}).
In similar way we find for the tensor channel
\begin{equation}
\langle N|\bar q\sigma_{ij}q|N\rangle=\sum\varepsilon_{ijk}\chi^{*}{\sigma_k}\chi\cdot\int d^3r[f_1^2(r)+\frac{2}{3}f_2^2(r)]\,.
\label{45}
\end{equation}

In particular calculations we employ the relativistic quark model developed in \cite{9}. In its version presented in \cite{8} the valence quarks are described by the functions $\psi_u=\psi_d=\psi$ with
\begin{equation}
\psi({\bf r})=N\exp{(-\frac{r^2}{2R^2})}\left( \begin{array}{c}
\chi \\
i\beta{\bfsig}\cdot{\bf x} \chi\end{array} \right); \quad {\bf x}=\frac{{\bf r}}{R},
\label{46}
\end{equation}
in agreement with Eq.~(\ref{43}). Parameters of the model $R=0.6$~fm and $\beta=0.39$ are fitted to reproduce the values of
the proton charge radius and of the axial coupling constant correspondingly. Factor $N$ is determined by the normalization condition $\int d^3r \psi({\bf r})\gamma_0\psi({\bf r})=1$ expressing the conservation of the baryon charge.
The treatment of sea quarks in the model suggested in \cite{9} is not important for us.

Direct calculations provide for the contribution of $4u$ operator in the $AA$ channel
\begin{equation}
\Pi^{3N}_{4q}=-\frac{13J_A^2}{36q^2}\hat q\rho^2+\frac{13J_A^2}{36q^2}\frac{(Pq)}{m^2}\hat P\rho^2,
\label{47}
\end{equation}
where
\begin{equation}
J_A=\frac{1-\beta^2/2}{1+3\beta^2/2}=0.75.
\label{48}
\end{equation}
The contribution to the left hand side of the sum rules presented by Eq.~(\ref{X4}) is thus
\begin{equation}
\tilde A_6^{3N}= \frac{13J_A^2}{36}\cdot 32\pi^4\rho^2; \quad \tilde C_6^{3N}=-\frac{13J_A^2}{36}\frac{s-m^2}{2m^2}\cdot 32\pi^4\rho^2.
\label{49}
\end{equation}
Contribution of $2u2d$ operator in the $AA$ channel is
\begin{equation}
\Pi^{3N}_{4q}=-\frac{2J_A^2}{q^2}\hat q\rho^2-\frac{J_A^2}{q^2}\frac{(Pq)}{m^2}\hat P\rho^2,
\label{50}
\end{equation}
providing
\begin{equation}
\tilde A_6^{3N}= 2J_A^2\cdot 32\pi^4\rho^2; \quad \tilde C_6^{3N}=J_A^2\frac{s-m^2}{2m^2}\cdot 32\pi^4\rho^2.
\label{51}
\end{equation}

Only the $4u$ operator contributes in the tensor $TT$ channel
\begin{equation}
\Pi^{3N}_{4q}=\frac{13J_T^2}{72q^2}\hat q\rho^2+\frac{13J_A^2}{18q^2}\frac{(Pq)}{m^2}\hat P\rho^2,
\label{52}
\end{equation}
with
\begin{equation}
J_T=\frac{1+\beta^2/2}{1+3\beta^2/2}=0.88.
\label{48t}
\end{equation}
This provided
\begin{equation}
\tilde A_6^{3N}= -\frac{13J_T^2}{72}\cdot 32\pi^4\rho^2; \quad  \tilde C_6^{3N}=-\frac{13J_T^2}{18}\frac{s-m^2}{2m^2}\cdot 32\pi^4\rho^2.
\label{54}
\end{equation}
The operator containing two $u$ quarks and two $d$ quarks contributes in the $AT$ channel providing
\begin{equation}
\Pi^{3N}_{4q}=\frac{2J_AJ_T}{q^2}\frac{(Pq)}{m}\rho^2,
\label{55}
\end{equation}
leading to the contribution
\begin{equation}
\tilde B_6^{3N}=-J_AJ_T\frac{s-m^2}{m}\cdot 32\pi^4\rho^2.
\label{56}
\end{equation}

\section{Solution of the sum rules equations}

Now we solve the sum rules equations given by Eqs.~(\ref{X3},\ref{X4}) including the contributions $\tilde B_3^{3N}$, $\tilde A_6^{3N}$, $\tilde B_6^{3N}$ and $\tilde C_6^{3N}$ obtained above. The values of the nucleon self energies depend on the value of the cutoff $\mu$ employed in calculations of contribution of the nonlinear scalar condensate. We found that at $\rho=\rho_0$ the $3N$ forces subtract $25$~MeV from the nucleon effective mass $m^*$ if we put the cutoff parameter $\mu=882$ MeV. This value changes between 45 MeV and 13 MeV while $\mu$ varies between 600 and 1200 MeV. The $3N$ forces subtract also $37$~MeV from $\Sigma_V$. The uncertainty of this value due to possible variation of $\mu$ is about 1 MeV.

Now we analyze how the values of the nucleon self energies at $\rho=\rho_0$ change after separate inclusion of the nonlinear scalar condensate (NSC) and of the four-quark condensates in various channels. The results are presented in Table 1. Note that the contribution of the four-quark condensate in the $PP$ channel vanishes. The separate contributions are not additive. This is mainly because the self energies are connected with the condensates by the nonlinear expressions given by Eq.(\ref{57}). Also the values for the continuum threshold $W_m$ are slightly different for different channels.

One can see that there are the contributions of the same order to the effective mass $m^*$ provided by the NSC  and by the four-quark condensates in $SS$, $VS$ and $AT$ channels. The contributions of the four-quark condensates cancel to large extent. Thus the contribution to $m^*$ of the $3N$ forces can be viewed as caused mainly by the nonlinear scalar condensate. The NSC does not enter the second equality of Eq.~(\ref{57}) and does not provide a noticeable contribution to the vector self energy $\Sigma_V$. The largest contribution to the latter is caused by the four-quark condensates in $SS$ and $VV$ channels.

\begin{table}
\begin{center}
\caption{Contributions of the nonlinear part of the scalar condensate (NSC) and of the four-quark condensates in various channels in configurations where the two pairs of quark operators act on two different in-medium nucleons to the nucleon self-energies $\delta m^*$ and $\delta \Sigma_V$($\rho=\rho_0$, $\mu=882$ MeV)}.
\begin{tabular}{|c|c|c|}
\hline
    &$\delta m^*$,MeV&$\delta \Sigma_V$, MeV\\
\hline
NSC&-26&-3\\
\hline
SS&-25&-11\\
\hline
VV&7&-16\\
\hline
VS&33&5\\
\hline
AA&-4&-6\\
\hline
TT&4&-5\\
\hline
AT&-14&-2\\
\hline
\end{tabular}
\end{center}
\end{table}

The density dependence of nucleon parameters is presented in Figs.~6 and 7. More detailed data in the vicinity of the density $\rho_0$ is given in Table 2. One can see that the $3N$ forces provide negative contributions to both the nucleon mass $m^*$ and the vector self energy $\Sigma_V$. Finally the $3N$ terms provide a negative contribution to the potential energy $U<0$ thus increasing the value of $|U|$. The values of the nucleon pole residue $\lambda_m^2$ and of the continuum threshold $W_m^2$ suffer small changes after inclusion of the $3N$ forces.

\begin{table}
\caption{Influence of the $3N$ forces on the density dependence of nucleon parameters. For each value of $\rho/\rho_0$
the upper and lower lines show the results with the $3N$ forces neglected and included correspondingly ($\mu=882$ MeV).}
\begin{center}
\begin{tabular}{|c|c|c|c|c|c|}
\hline
$\rho/\rho_0$&$m^*$,MeV&$\Sigma_V$, MeV&$U$, MeV&$\lambda_m^2$,GeV$^6$&$W_m^2$, GeV$^2$\\
\hline
0.90&640&196&-93&1.43&1.75\\
&618&166&-145&1.44&1.79\\
\hline
0.95&620&209&-100&1.39&1.73\\
&596&176&-157&1.41&1.79\\
\hline
1.00&599&222&-108&1.36&1.72\\
&574&185&-169&1.39&1.79\\
\hline
1.05&577&236&-115&1.32&1.71\\
&551&-195&-183&1.37&1.79\\
\hline
1.10&555&250&-123&1.29&1.70\\
&528&204&-196&1.36&1.80\\
\hline

\end{tabular}
\end{center}
\end{table}

We can make some predictions on the contributions of the $4N$ forces to the nucleon self energies. To obtain these contributions we must include the $3N$ forces between the nucleons of the matter in the term $\tilde B_3$ which is the ingredient of the function ${\cal L}^I(M^2, W_m^2)$ in the first equality of Eq.~(\ref{57}). This means that we need the value of $\sigma_N^{eff}$ with inclusion of the $3N$ forces of the matter to obtain $\tilde B_3$. The function $\sigma_N^{eff}$ with inclusion of the $2N$ and the $3N$ interactions between the nucleons of the matter is found in \cite{7}. We also need the scalar-scalar quark condensate in configuration where two pairs of quarks act on two different nucleons of the matter, since this condensate also depends on $\sigma_N^{eff}$. This provides the term $\tilde A_6$ which enters the function ${\cal L}^q(M^2, W_m^2)$.
 Following our analysis, in this term we need the function $\sigma_N^{eff}$ which includes only the $2N$ interactions between the nucleons of the matter.

The function $\sigma_N^{eff}$ with inclusion of $2N$ interactions was obtained in the present paper. It was demonstrated that $\sigma_N^{eff}-\sigma_N \ll \sigma_N$. Thus we expect the influence of the $3N$ forces between the nucleons of the matter on ${\cal L}^I(M^2, W_m^2)$  to be small. On the other hand, one can see from Fig.~5 of the paper \cite{7} that $\sigma_N^{eff}-\sigma_N \approx -15$ MeV at the saturation value of nuclear density if $2N$ and $3N$ forces between the nucleons of the matter are included. One can see from Eq.~(\ref{57}) that the difference $m^*-m$ is approximately proportional to the value of $\sigma_N^{eff}$. Since $m^*-m =350$ MeV at $\sigma_N=45$ MeV, we expect the $4N$ forces to add about 100 MeV to the effective mass $m^*$. The influence of the $4N$ forces on the vector self energy $\Sigma_V$ is expected to be small.

\section{Summary}

We obtained the contribution of the $3N$ forces to the nucleon characteristics in symmetric nuclear matter. The most important contributions are due to the nonlinear scalar condensate and to the four-quark condensates in the scalar-scalar and
 vector-vector channels with two diquark operators acting on two different nucleons of the matter. They are shown in the diagrams of Fig.~2. The contributions are either expressed in terms of pion-nucleon sigma-term $\sigma_N$ or calculated in a model-independent way.

We found that the $4N$ forces are expected to add about 100 MeV to the proton effective mass $m^*$.  They are expected to provide only small contribution to the vector self energy $\Sigma_V$. However, the detailed analysis of the contribution of $4N$ forces is beyond the scope of the present paper. It is rather the subject of future work.
\\
\section*{Acknowledgement}
We acknowledge  support by the RSF grant 14-22-00281.

{}

\section{Figure captions}

\noindent
Fig.~1. The Feynman diagrams providing the main contribution to the nucleon parameters in the gas approximation.
The helix line shows the polarization operator. Solid lines denote the quarks. The bold lines are for the nucleons of nuclear matter.\\

\noindent
Fig.~2. The Feynman diagrams providing the main contribution to the nucleon parameters corresponding to inclusion of the $3N$ forces. Dashed block denotes the pion field. Other notations are the same as in Fig.~1.\\

\noindent
Fig.~3. The Feynman diagrams of nucleon-nucleon interactions corresponding to our approach.Only the $2N$ forces are included. Solid lines are for the probe nucleon, bold lines denote the nucleons of the matter. The dashed and wavy lines are for the scalar and vector mesons correspondingly. Dotted and dashed-dotted lines are for the mesons with combinations of the quantum numbers $SS$, $VV$, $PP$, $AA$, $TT$, $VS$ and $AT$.\\

\noindent
Fig.~4. Additional Feynman diagrams of nucleon-nucleon interactions corresponding to inclusion of the $3N$ forces.
The shaded block denotes the pion field. Other notations are the same as in Fig.~3. The two nucleons of the matter are in the same space-time point $x$.

\noindent
Fig.~5. Density dependence of the effective pion-nucleon sigma term which includes the $2N$ interactions between the nucleons of the matter. Dashed, solid and dotted lines are for $\mu=600$, 882 and 1200 MeV correspondingly.\\

\noindent
Fig.~6. Density dependence of the effective nucleon mass $m^*$, of the vector self energy $\Sigma_V$ and of the single-particle potential energy $U$. The horizontal axis corresponds to the density $\rho$ related to its saturation value $\rho_0$. The vertical axis is for $m^*$, $\Sigma_V$ and $U$. The dashed and solid lines show the results with the $3N$ forces omitted and included correspondingly.\\

\noindent
Fig.~7. Density dependence of the nucleon residue $\lambda_m^2$ and of the continuum threshold $W_m^2$.
The notations are the same as in Fig.~6.

\newpage

\begin{figure}
\centerline{\epsfig{file= 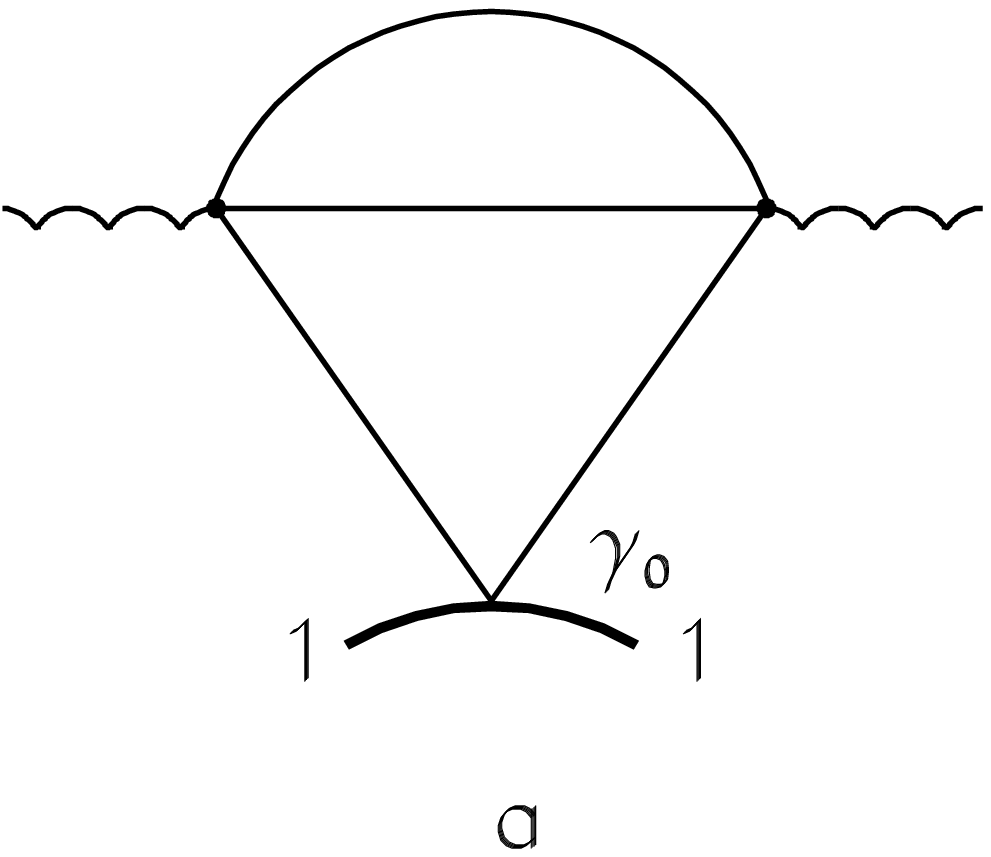,width=6.0cm}
\epsfig{file= 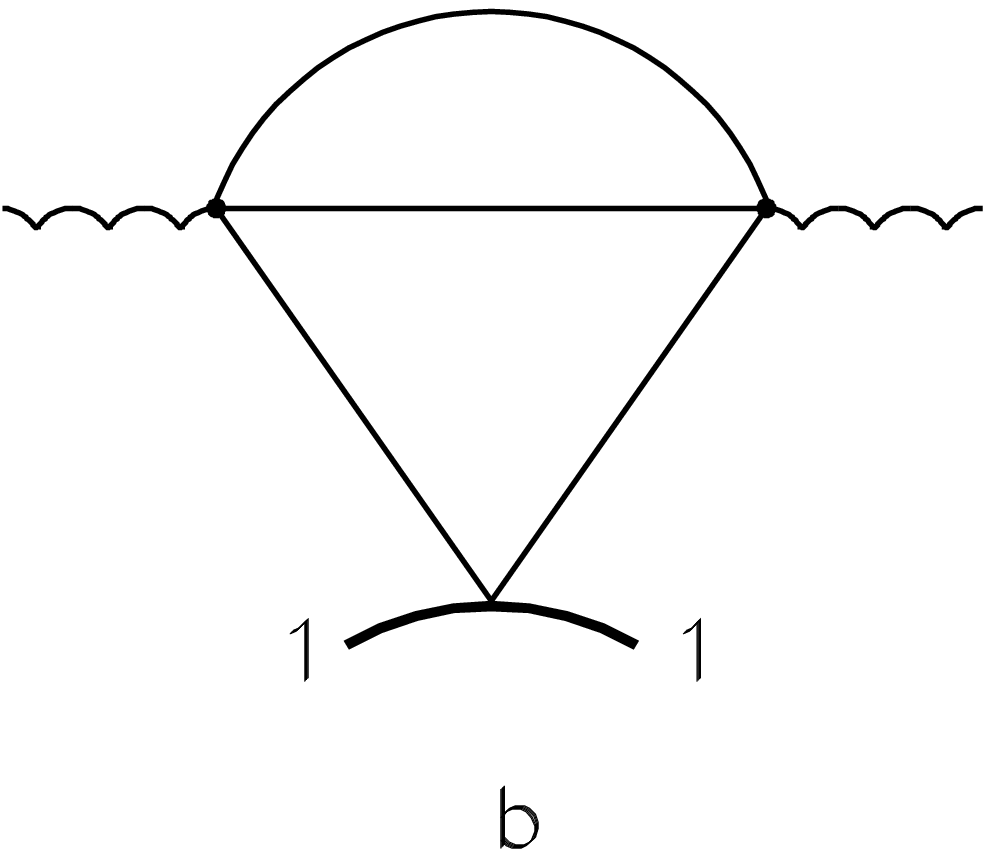,width=6.0cm}}
\centerline{\epsfig{file= 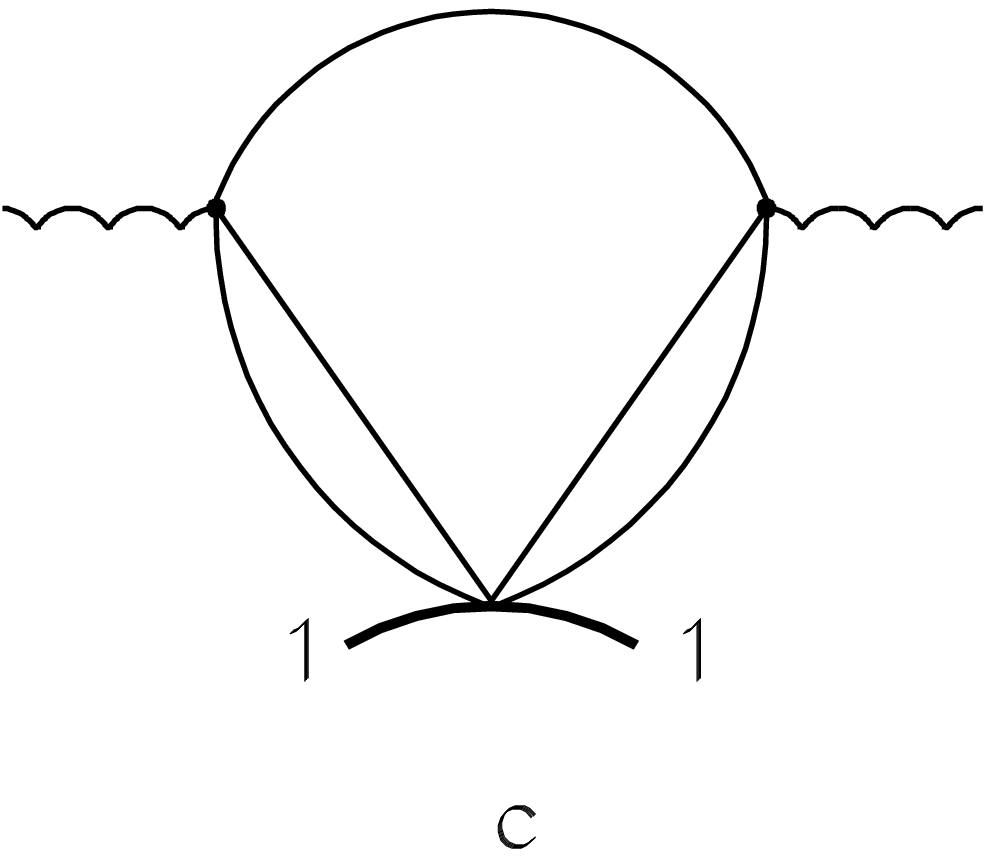,width=6.0cm}}
\caption{}
\end{figure}

\begin{figure}
\centerline{\epsfig{file= 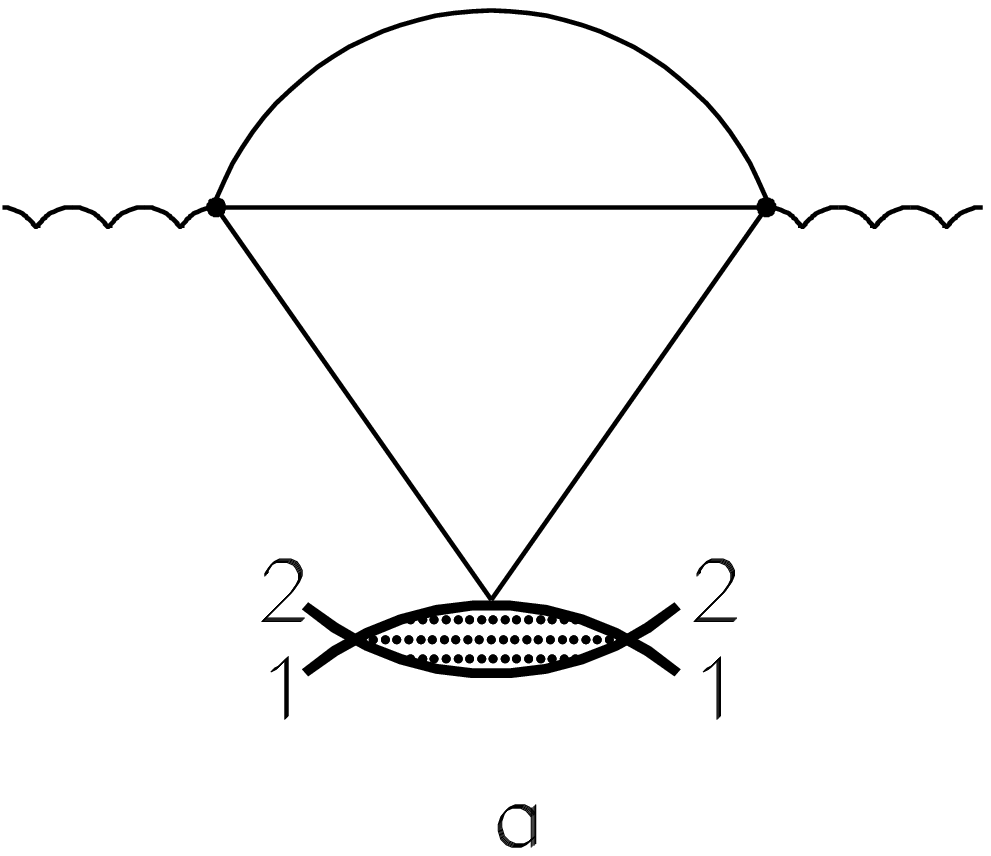,width=6.0cm}
\epsfig{file= 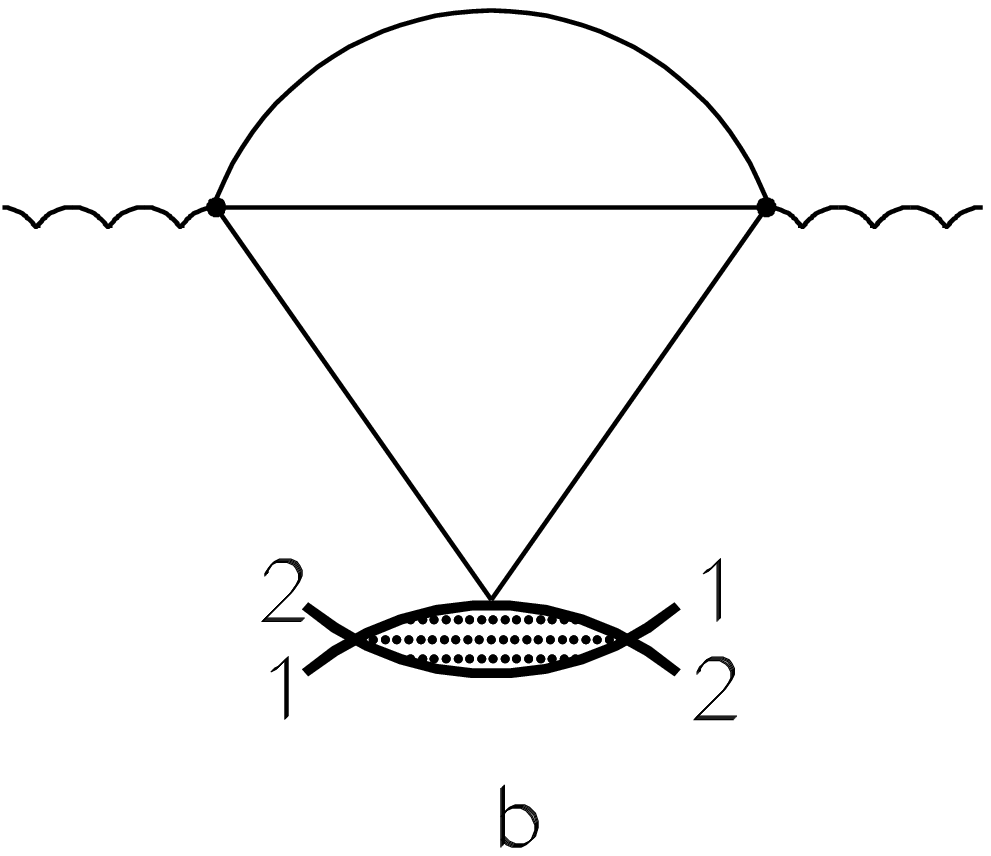,width=6.0cm}}
\centerline{\epsfig{file= 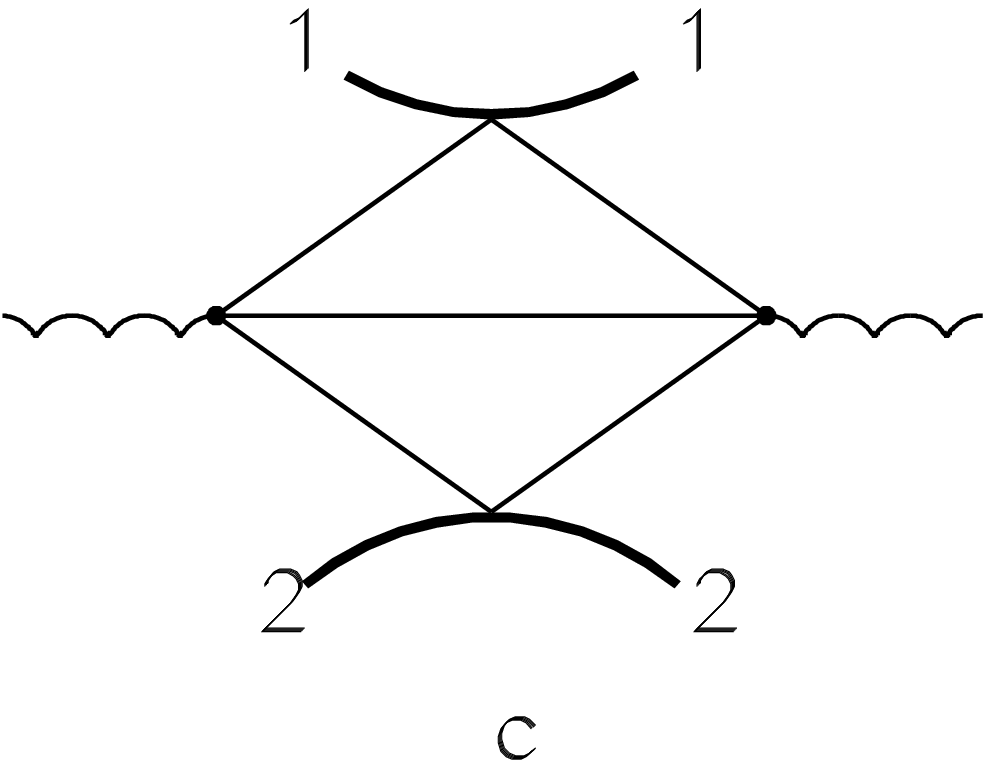,width=6.0cm}}
\caption{}
\end{figure}

\begin{figure}
\centerline{\epsfig{file= 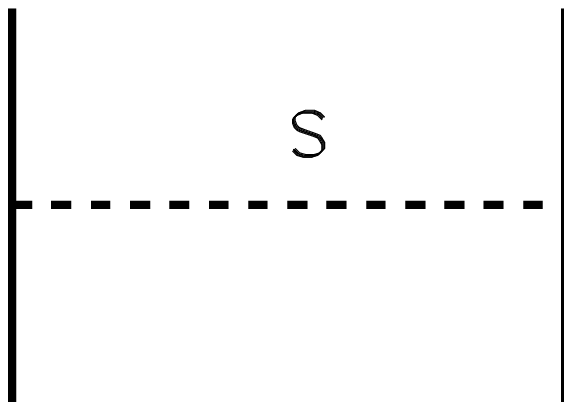,width=6.5cm} \hspace{-2cm} \epsfig{file= 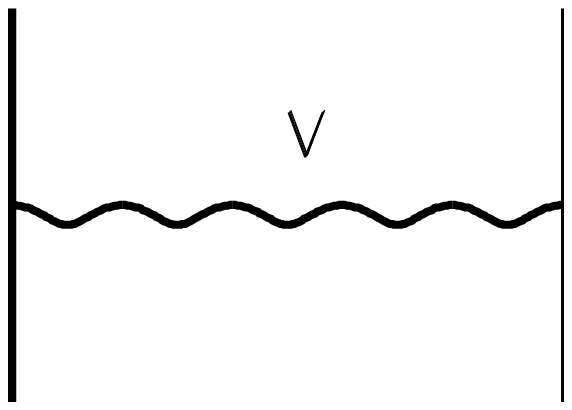,width=6.5cm} \hspace{-2cm} \epsfig{file= 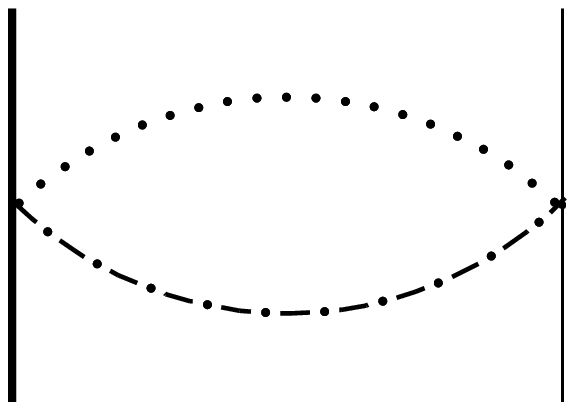,width=6.5cm}}
\caption{ }
\end{figure}
\begin{figure}
\centerline{\epsfig{file= 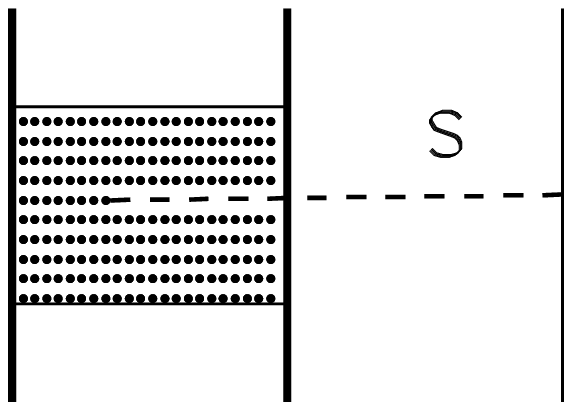,width=7cm} \epsfig{file= 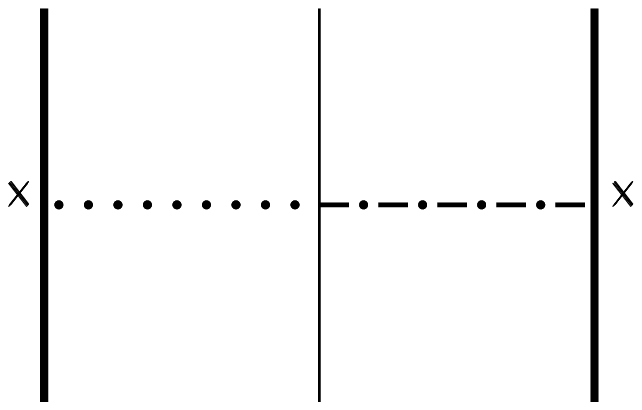,width=7cm}}
\caption{}
\end{figure}

\begin{figure}
\centerline{\epsfig{file= 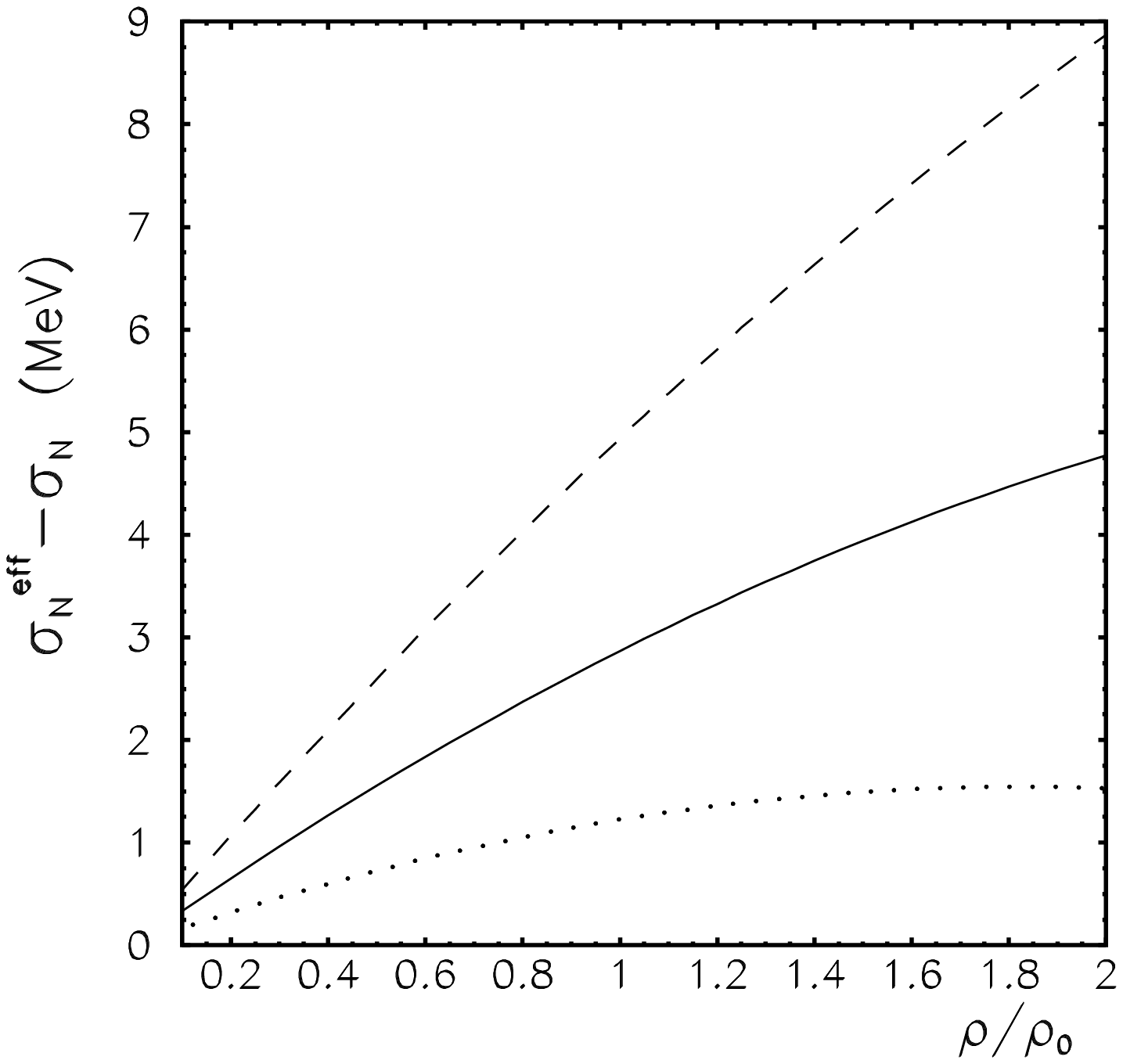,width=9.0cm}}
\caption{}
\end{figure}

\begin{figure}
\centerline{\epsfig{file= 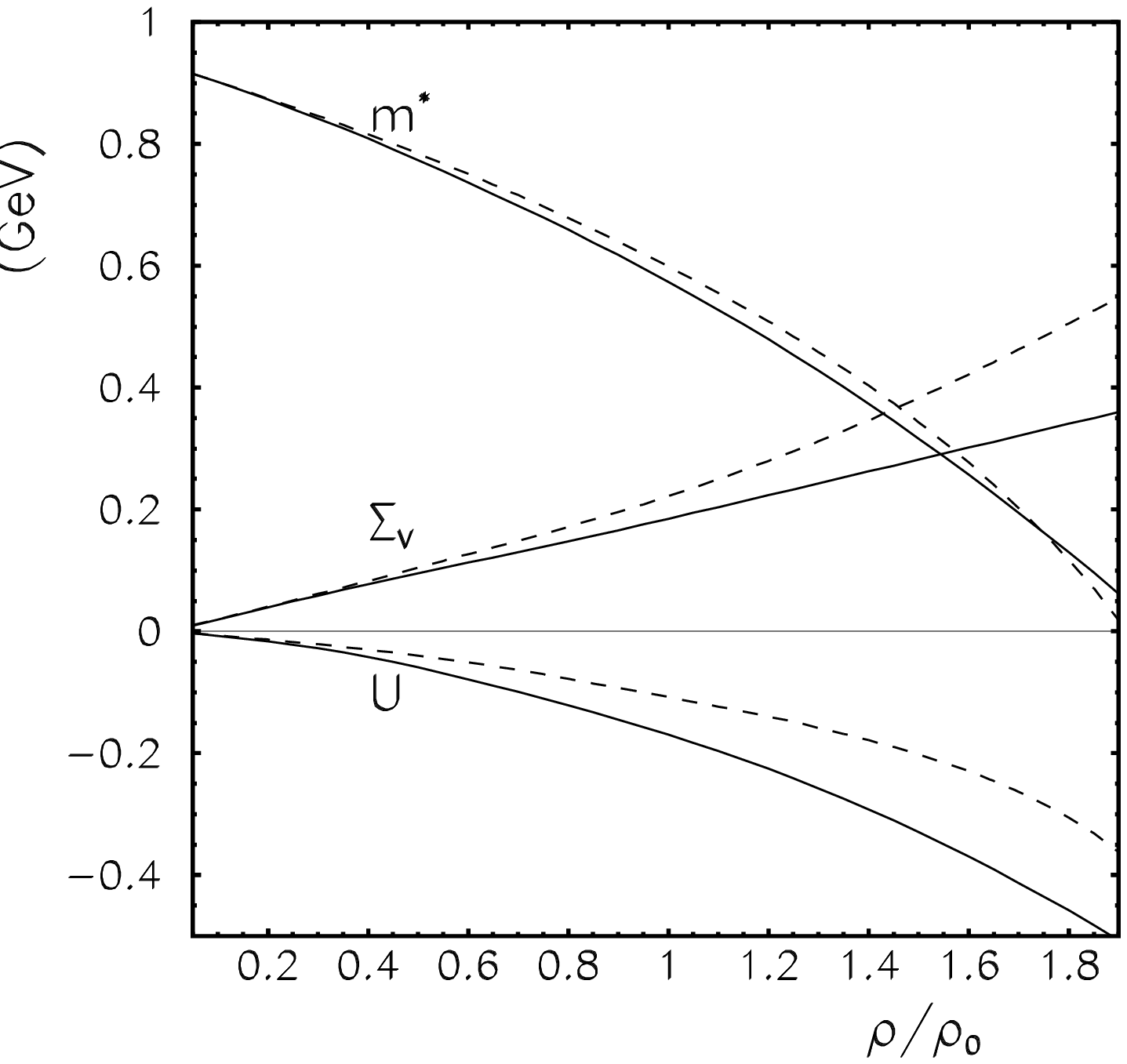,width=9.0cm}}
\caption{}
\end{figure}

\begin{figure}
\centerline{\epsfig{file= 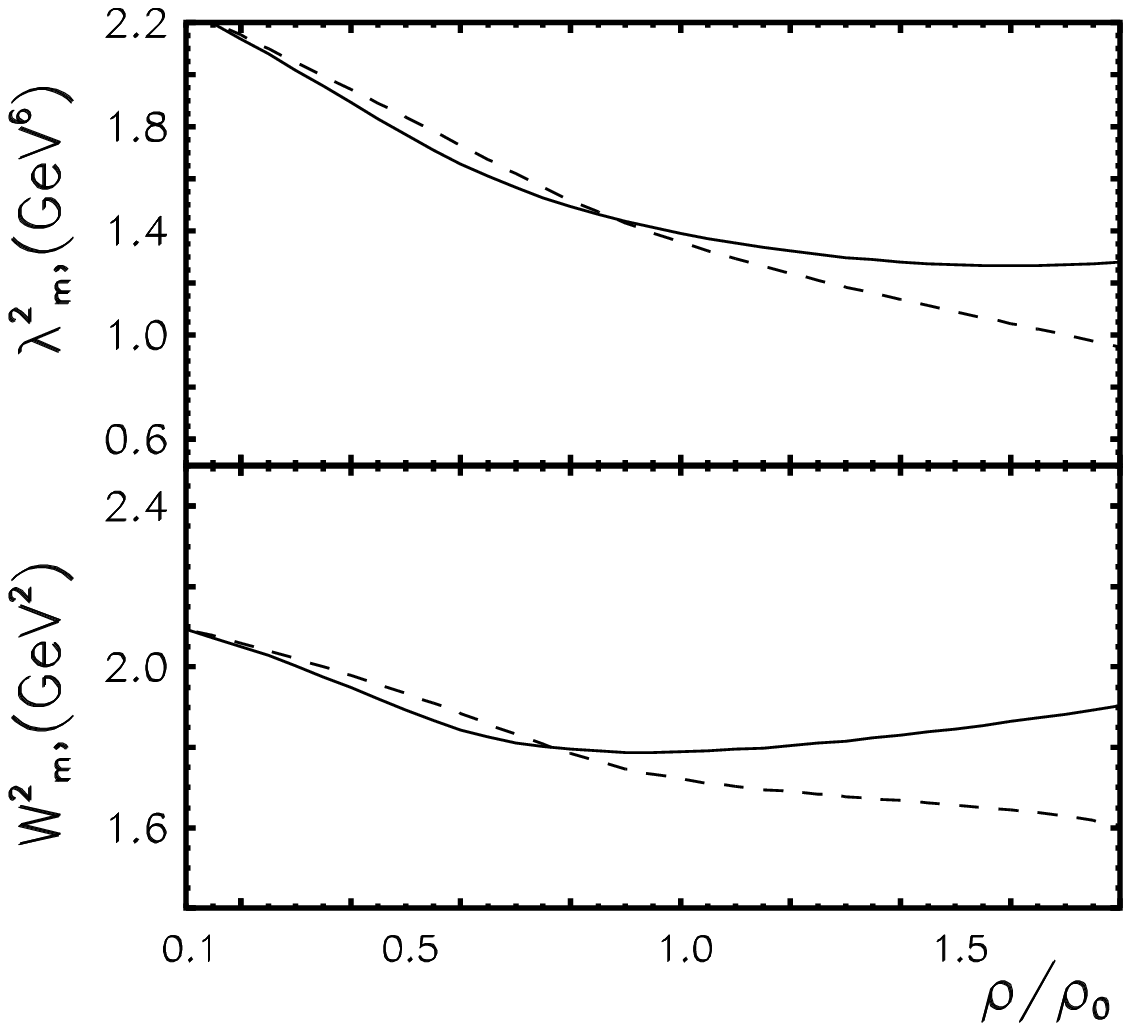,width=10.0cm}}
\caption{}
\end{figure}

\end{document}